\documentclass[conference]{IEEEtran}

\usepackage[pdftex]{graphicx}
\usepackage{cite, balance}
\usepackage{fixltx2e}
\usepackage{epstopdf}
\usepackage{amssymb}
\usepackage[cmex10]{amsmath}
\usepackage{multirow}
\usepackage{algorithmic}
\usepackage{algorithm}
\usepackage[utf8]{inputenc}
\usepackage{soul, color}
\usepackage[caption=false,font=footnotesize]{subfig}

\begin{document}


\title{Frequency Switching for Simultaneous Wireless Information and Power Transfer}

\author{

\IEEEauthorblockN{Dogay Altinel\IEEEauthorrefmark{1} \IEEEauthorrefmark{7}, Gunes Karabulut Kurt\IEEEauthorrefmark{1}}
	
\IEEEauthorblockA{\IEEEauthorrefmark{1}Department of Electronics and Communication Engineering, Istanbul Technical University, Turkey}
   
\IEEEauthorblockA{\IEEEauthorrefmark{7}Department of Electrical and Electronics Engineering, Istanbul Medeniyet University, Turkey
\\\ dogay.altinel@medeniyet.edu.tr, \{altineld, gkurt\}@itu.edu.tr}
	
}

\maketitle

\maketitle

\textcolor{blue}{}
\begin{abstract} 
A new frequency switching receiver structure is proposed  for simultaneous wireless information and power transfer in multi-carrier communication systems.  Each subcarrier is switched to either the energy harvesting unit or the information decoding unit, according to the optimal subcarrier allocation. To implement the system, one-bit feedback is required for each subcarrier. Two optimization problems are defined, converted to binary knapsack problems, and solved using dynamic programming approaches. Upper bounds are obtained using continuous relaxations. Power allocation is integrated to further increase the performance. Numerical studies show that the proposed frequency switching based model is better than existing models in a wide range of parameters. 
%
%
\end{abstract}



\vspace{0.2cm}
\section{Introduction}
\label{sec:introduction}
 


Simultaneous wireless information and power transfer (SWIPT) concept, which is recently developed  in the literature, integrates wireless power transfer to communication technologies \cite{4595260}. 
The main goal of SWIPT is to provide energy to the receiver by means of radio frequency (RF) signals captured from the transmitter, while transmitting information.  Actually, the process of decoding information and harvesting energy from the same RF signal simultaneously is not possible for practical circuits. In \cite{6489506},  time switching (TS) and power splitting (PS) based practical receiver designs are proposed for the co-located receivers. 
In TS designs, the receiver antenna switches between the energy harvester and the information decoder according to a time schedule. On the other hand, in PS designs, the received radio signal is divided into two signals with desired powers for the energy harvester and the information decoder. In \cite{6623062}, dynamic PS operation scheme is proposed for separated and integrated receiver architectures. 

In multi-carrier communication (MCC), the available bandwidth is divided into a number of narrowband subcarriers to obtain flat channels and high data rate. Based on narrowband subcarrier structure, MCC is an appropriate technique for SWIPT systems. 
In this context, in \cite{6589954}, a dual-antenna mobile architecture and a framework are presented for realizing SWIPT in broadband wireless systems. In \cite{7247664}, an algorithm is proposed to optimize TS and power allocation jointly for multi-carrier relay network. While these works exploit frequency diversity in multi-carrier based approaches to improve the efficiency, 
they use PS or TS techniques to share the received signal between the information decoder and the energy harvester. In \cite{7378853}, in order to exploit frequency diversity, the subcarrier seperation scheme is proposed  in a multiuser OFDM system. Although the transfer of information and power on different subcarriers is adopted in \cite{7378853}, the structure and implementation model of receiver is not given. Additionally, the maximization of total harvested power under the channel capacity constraint is not considered.

\begin{figure}[t]
\centering
\includegraphics[trim=3.3cm 0cm 3.7cm 0cm, width=0.32\textwidth]{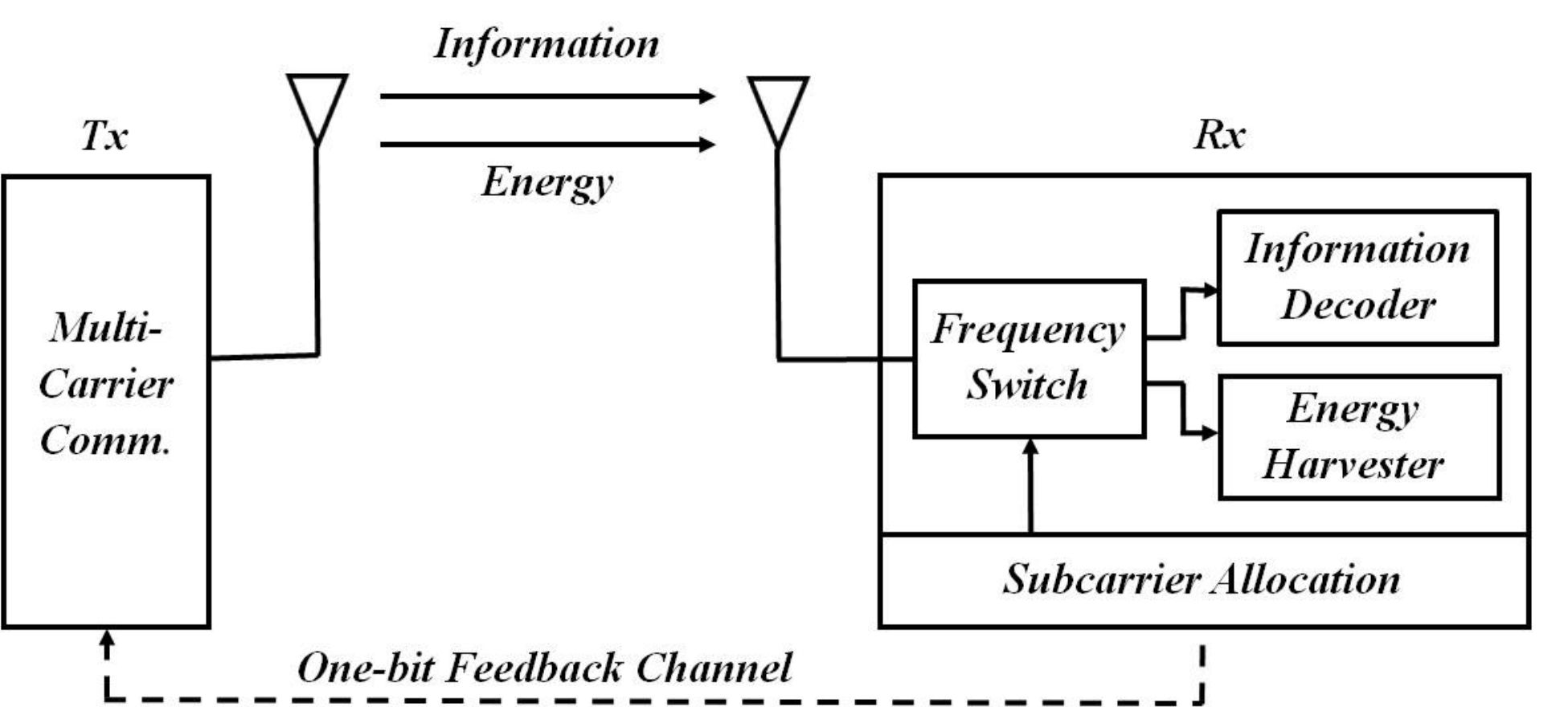}
\caption{System model for simultaneous wireless information and power transfer with FS receiver.}
\label{fig:systemmodel1}
\end{figure}

In this paper, we propose a new receiver structure making use of frequency switch (FS) to improve the performance of SWIPT in MCC systems, using one-bit feedback per subcarrier. We define two different optimization problems, based on the power and  channel capacity requirements of the system, and convert them to binary knapsack problems that can be addressed using dynamic programming approaches. We also provide upper bounds for the achievable information and power transfer limits. We then optimize transmit power levels to further improve the performance. Numerical results are provided and compared with TS and PS based receivers.

\vspace{0.2cm}
\section{System Model and Receiver Structure}

The considered MCC system consists of a transmitter and a receiver, as illustrated in Fig. \ref{fig:systemmodel1}. The transmitter sends both information and power in the same multi-carrier symbol to the receiver that is equipped with both an information decoder and an energy harvester. Assuming $K$ subcarriers that are non-overlapping in the frequency spectrum, the  received symbol on the $k^ \textrm{th}$ subcarrier can be modeled as \begin{eqnarray}
Y_k = \sqrt{P_{t,k}} H_k X_k + Z_k,  \;k = 1, 2,\ldots, K.
\label{channelmodel}
\end{eqnarray}
Here, $X_k$ represents the transmitted symbol on the $k^\textrm{th}$ subcarrier. $Z_k$ is the complex additive white Gaussian noise (AWGN) with zero mean and $\sigma^2_z$ variance. $H_k$ is the channel coefficient affecting the transmitted symbol between the transmitter and the receiver. We denote the average transmit power on each subcarrier by $P_{t,k}$, and assume that   $E[|X_k|^2] = 1$, $\forall k$. $E[\cdot]$ represents the expectation operator. 

\begin{figure}[t]
\centering
\includegraphics[trim=3.5cm 0cm 3.5cm 0cm, width=0.15\textwidth]{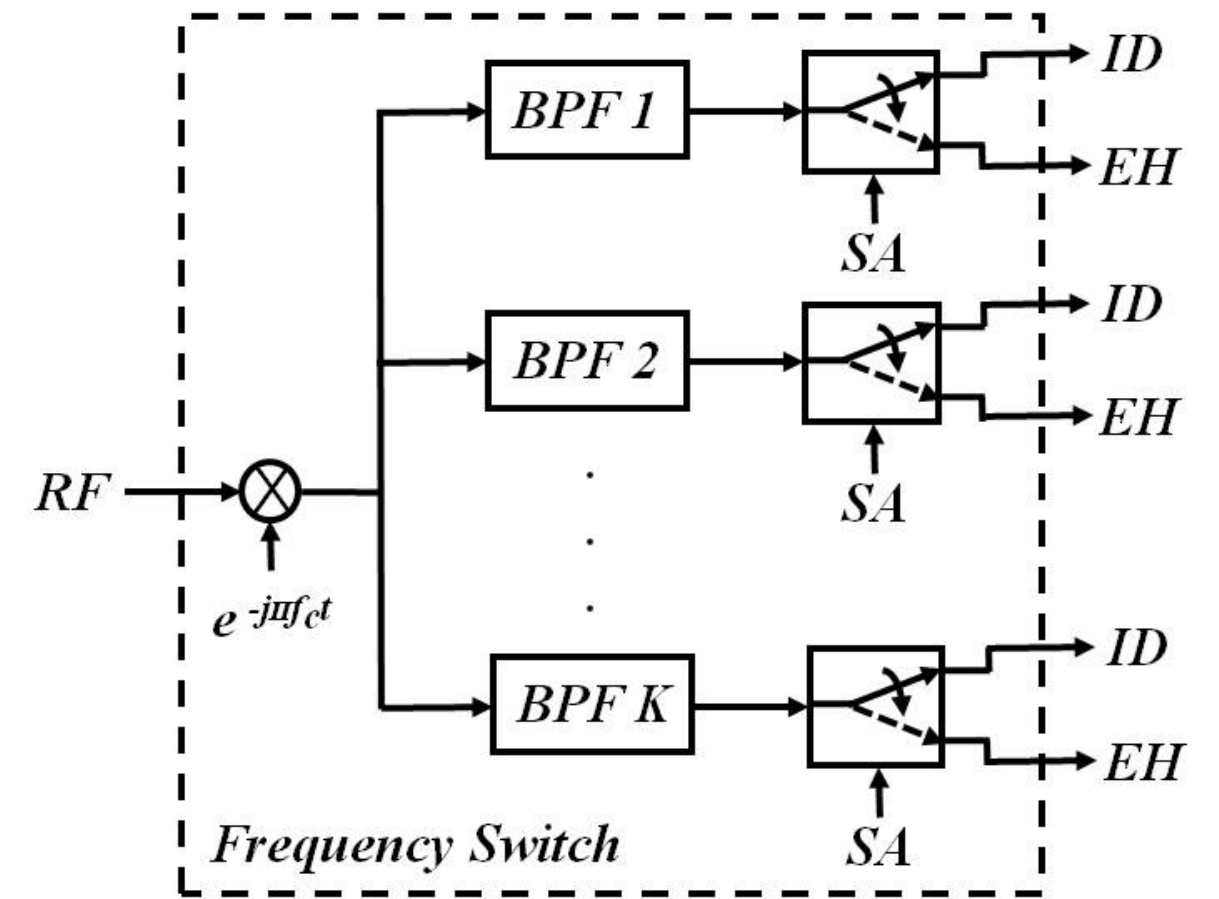}
\caption{Internal structure of the proposed frequency switch. Band pass filter for each subcarrier is denoted by BPF $k$, $k=1,2,\ldots,K$. Here, SA, ID, and EH represent subcarrier allocation, information decoder, and energy harvester, respectively.}
\label{fig:systemmodel2}
\end{figure}

In the proposed FS receiver, we target to utilize each subcarrier for either information or power transfer.  The internal structure of an examplary frequency switch is shown in Fig.\ref{fig:systemmodel2}. 
This can be realized as a 
filterbank \cite{995073, 5753092}. Since it is proven that energy harvesting is possible in baseband \cite{6623062, Cid-FuentesNBCCA16-2}, the frequency switch first performs baseband conversion, 
then, after band pass filtering it performs switching of subcarriers to the information decoder or the energy harvester dynamically, depending on the subcarrier allocation (SA) scheme. 
The proposed receiver can be used in filterbank multi-carrier communication (FBMC) and orthogonal frequency division multiplexing (OFDM) systems.


In order to optimally allocate subcarriers to information and power transfer functionalities, we need to consider the channel capacity and the harvested power. To identify the subcarrier set to be used for information transfer, the channel capacity  $(C_k)$ of the $k^ \textrm{th}$ subcarrier can be obtained as 
\begin{eqnarray}
C_k=B \log_2(1+|H_k|^2 \gamma_k),  
\label{capacity1}
\end{eqnarray}
where $B$ is the subcarrier bandwidth and  $\gamma_k$ represents the average signal to noise ratio,  expressed as $\gamma_k=P_{t,k}/\sigma^2_z$. 
The channel capacity is proportional to the square of channel coefficient, $C_k \propto |H_k|^2 \gamma_k$, in dB scale.

On the other hand, to determine the subcarrier set that will be used for power transfer, we can consider the relationship between the channel coefficient and the harvested power from $k^\textrm{th}$ subcarrier $(Q_{k})$ expressed as \cite{6623062}
\begin{eqnarray}
Q_k=\eta_k |H_k|^2 P_{t,k},  
\label{harvestedpower1}
\end{eqnarray}
where $\eta_k$ is the conversion efficiency of RF signal to direct current signal for each subcarrier channel. The value of $\eta_k$  mainly depends on the design of the energy harvester, which can be around 0.5 in commercial applications \cite{1113333}.  
It can be seen that the harvested power is also proportional to the square of channel coefficient, $Q_{k} \propto |H_k|^2 \gamma_k$, in linear scale.

In the proposed system, it is assumed that the channel coefficients are available at the receiver. On the receiver side, we define an optimization problem and run a SA algorithm. The outcome of the algorithm, indicating the functionality of each subcarrier, is fed to both the frequency switch and to the  transmitter using a feedback channel, requiring one-bit information for each subcarrier. Let $\mathcal{S}$ represents the set of subcarriers to be used for information transfer. Then, $\mathcal{S}^c =  \{1,2,\ldots,K\}\setminus \mathcal{S}$ becomes the subcarrier set to be used for power transfer. The corresponding total channel capacity $(C_T)$ becomes 
\begin{eqnarray}
C_T= \sum\limits_{k \in \mathcal{S}} B \log_2(1+|H_k|^2 \gamma_k),
\label{capacity2}
\end{eqnarray}
and the total harvested power $(Q_T)$ can be calculated as
\begin{eqnarray}
Q_T= \sum\limits_{k\in \mathcal{S}^c} \eta_k |H_k|^2 P_{t,k}. 
\label{harvestedpower2}
\end{eqnarray}
There is a clear tradeoff in the selection of $\mathcal{S}$ and $\mathcal{S}^c$ in terms of the total channel capacity versus the total harvested power. The channel coefficients and the transmit powers have an important role on both quantities. Based on these two parameters, it can be concluded that  the performance of the SWIPT system can be increased in terms of either the channel capacity or the harvested power, as will be investigated next. 





\section{Subcarrier Allocation}

The SA problem, i.e., selection of  $\mathcal{S}$ (or equivalently $\mathcal{S}^c$), can significantly affect the performance of the SWIPT system and can be formulated as an optimization problem. Here, it is assumed that the transmit powers are equal, $P_{t,k}=P_{t,e}$, $\forall k$. We concentrate on two optimization problems: maximizing the total channel capacity  while harvesting a desired amount of power (\textbf{P1}), and  maximizing the total harvested power while satisfying a minimum channel  capacity (\textbf{P2}). 
Let $s_k \in \{0,1\}$ be an indicator function, $\forall k$.  $s_k=1$ indicates that the subcarrier  is allocated for the transmission of information. Otherwise, i.e. $s_k=0$, the subcarrier is used for power transmission. $s_k^c$, an indicator function for the subcarrier to be harvested, can be obtained as $s_k^c=1-s_k$. We can define vectors of these indicator fuctions as $\textbf{s} =[s_1,s_2,\ldots,s_K]$ and 
$\textbf{s}^c =[s_1^c,s_2^c,\ldots,s_K^c]$. Note that  $\textbf{s}^c = \textbf{1} - \textbf{s}$, where $\textbf{1}$ is a vector of ones of length $K$. The indices of ones in $\textbf{s}$ ($\textbf{s}^c$)  consitute $\mathcal{S}$ ($\mathcal{S}^c$), respectively. 

The first optimization problem, (\textbf{P1}), maximizing the total channel capacity $(C_T)$ while harvesting a minimum power $(Q_{min})$ can now be stated as 
\begin{eqnarray}
(\textbf{P1}): \quad
\begin{aligned}
& \underset{\textbf{s} }{\text{max}}
& & \sum\limits_{k=1}^K s_k C_k \\
& \text{s.t.}
& & \sum\limits_{k=1}^K s^{c}_{k} Q_k \geq Q_{min}, \\
&&& s_k,  s^{c}_{k} \in \{0,1\}, \\
&&& s_k^c=1-s_k.
\end{aligned}
\end{eqnarray}
The second optimization problem, (\textbf{P2}), maximizing the total harvested power $(Q_T)$ while guaranteeing a minimum capacity $(C_{min})$  can be formulated as
\begin{eqnarray}
(\textbf{P2}): \quad
\begin{aligned}
& \underset{\textbf{s}^c}{\text{max}}
& & \sum\limits_{k=1}^K s^{c}_{k}  Q_k \\
& \text{s.t.}
& & \sum\limits_{k=1}^K s_k C_k \geq C_{min}, \\
&&& s_k,  s^{c}_{k} \in \{0,1\}, \\ 
&&& s_k^c=1-s_k.
\end{aligned}
\end{eqnarray}

\vspace{0.1cm}
\subsection{Dynamic Programming for Optimal Solutions}

Dynamic programming breaks the problem down into smaller problems, and reuses the solution of small problems stored in the memory to find the optimal solution of the main optimization problem \cite{Bellman:1957}.  In case of SA problems in  (\textbf{P1}) and  (\textbf{P2}) dynamic programming can be used to obtain solutions.

Let the parameters of $Q_{th}=\sum_{k=1}^K Q_k - Q_{min}$ and $C_{th}=\sum_{k=1}^K C_k - C_{min}$ respectively denote the channel capacity threshold and the harvested power threshold. (\textbf{P1}) can be converted to
\begin{eqnarray}
(\textbf{P1}): \quad
\begin{aligned}
& \underset{\textbf{s}}{\text{max}}
& & \sum\limits_{k=1}^K s_k C_k \\
& \text{s.t.}
& & \sum\limits_{k=1}^K s_k Q_k \leq  Q_{th}, \\
&&& s_k \in \{0,1\},
\end{aligned}
\label{opt1}
\end{eqnarray}
and  (\textbf{P2}) can be written as
\begin{eqnarray}
(\textbf{P2}): \quad
\begin{aligned}
& \underset{\textbf{s}^c}{\text{max}}
& & \sum\limits_{k=1}^K s^{c}_{k}  Q_k \\
& \text{s.t.}
& & \sum\limits_{k=1}^K s^{c}_{k} C_k \leq C_{th}, \\ 
&&& s^{c}_{k} \in \{0,1\}.
\end{aligned}
\label{opt2}
\end{eqnarray}
We can clearly observe that (\ref{opt1}) and (\ref{opt2}) become a type of the binary knapsack problem, a well-known discrete programming problem \cite{martello1990knapsack}.  
Note that in the binary knapsack problem, $Q_k$, $C_k$,  $Q_{th}$, and $C_{th}$ are positive integers. However, in our case, although positive, they may not always be integers. To overcome this limitation, we scale all values with a proper factor.  We also assume that the threshold values in (\ref{opt1}) and (\ref{opt2}) are set as $Q_k \leq  Q_{th}$ and 
$C_k \leq  C_{th}$, $\forall k$. Now, the exact solution of  (\textbf{P1}) and  (\textbf{P2}) can be obtained using algorithms based on dynamic programming and branch-and-bound approach \cite{horowitz1974computing}. 
To determine the optimal solution with dynamic programming,  smaller problems can be formulated using iterations, and a recursive formulation for the $C_T$ ($Q_T$) can be obtained \cite{nemhauser1969discrete}.  
Finding the optimum solutions is also possible with the brute-force approach at the expense of high computational complexity. There are $2^K$ possible subsets of carriers, so the complexity becomes $O(2^K)$, where $O(\cdot)$ is complexity notation. However, the complexity is reduced to $O(K Q_{th})$ (for integer values of $Q_{th}$) in  (\textbf{P1}) with dynamic programming.

\vspace{0.1cm}
\subsection{Performance Bounds}

Although a closed form expression is not available for  (\ref{opt1}) or (\ref{opt2}), we can obtain an upper bound $(C_{up}$ or $Q_{up})$ that ensures $C_T \leq C_{up}$ or $Q_T \leq Q_{up}$ by utilizing continuous relaxation of the binary knapsack problem \cite{dantzig1957discrete}. For (\textbf{P1}), the ratios of the channel capacity to the harvested power can be ordered in a decreasing manner as 
\begin{eqnarray}
\frac{C_{l_1}}{Q_{l_1}} \geq  \frac{C_{l_2}}{Q_{l_2}} \geq \cdots \geq \frac{C_{l_K}}{Q_{l_K}},
\label{constraint2}
\end{eqnarray}
where $l_i \in\{1,\ldots,K\}$. Then,  starting with subcarrier index corresponding to the largest ratio $(l_1)$, the subcarrier channels are chosen for the transmission of information,
until the critical subcarrier  $(l_{d_1})$. The critical subcarrier is determined as the first subcarrier that exceeds the harvested power threshold, and its index is formally stated as 
\begin{eqnarray}
d_1={\text{min}} \left\{  d: \sum\limits_{i=1}^{d} Q_{l_i} > Q_{th} \right\},
\label{criticalchannel1}
\end{eqnarray}
where $d \in\{1,\ldots,K\}$.  In case of continuous relaxation, the  optimal solution for $s_{l_i}$ $(\hat{s}_{l_i})$ can be expressed as
\begin{eqnarray}
\hat{s}_{l_i}=
\begin{cases} 
1,& i=1,\ldots,d_1-1 \\ \frac{1}{{Q_{l_{d_1}}}}{ \sum\limits_{i=d_1}^K Q_{l_i} - Q_{min} },& i=d_1 \\ 0,& i=d_1+1,\ldots,K.
\end{cases}
\label{solutionS}
\end{eqnarray}
Hence, the upper bound of the total channel capacity becomes 
\begin{eqnarray}
C_{up}=\sum\limits_{i=1}^{d_1-1} C_{l_i} + \frac{C_{l_{d_1}}}{Q_{l_{d_1}}}  \left( \sum\limits_{i=d_1}^K Q_{l_i} - Q_{min} \right).
\label{upperbound1}
\end{eqnarray}
Note that the optimal solution of the continuous knapsack problem is considered as the upper bound for the optimization problem  (\textbf{P1}), where the  integer constraint for $s_{l_i}$ is relaxed.

Applying the same procedure  for (\textbf{P2}),  by relaxing the integer constraint for $s_{l_i}^c$, the upper bound of the total harvested power can be obtained as 
\begin{eqnarray}
Q_{up}=\sum\limits_{i=d_2+1}^{K} Q_{l_i} + \frac{Q_{l_{d_2}}}{C_{l_{d_2}}}  \left( \sum\limits_{i=1}^{d_2} C_{l_i} - C_{min} \right)
\label{upperbound2}
\end{eqnarray}
for 
\begin{eqnarray}
d_2={\text{min}} \left\{  d: \sum\limits_{i=d}^{K} C_{l_i} > C_{th} \right\},
\label{criticalchannel2}
\end{eqnarray}
where $d_2$ is the index of the critical subcarrier $l_{d_2}$ for (\textbf{P2}) according to the order in (\ref{constraint2}).

\vspace{0.1cm}
\section{Power Allocation}

In addition to the SA, it is possible to further increase the performance of the system model with power allocation (PA). We optimize each $P_{t,k}$ value for given  subcarrier sets according to the defined optimization problems. This scheme is referred to as 
subcarrier and power allocation (SPA). For  (\textbf{P1}) and (\textbf{P2}), the optimization problems are expressed as 
\begin{eqnarray}
(\textbf{P3}): \quad
\begin{aligned}
& \underset{\mathcal{P}_\textbf{t,c}}{\text{max}}
& & C_T \\ 
& \text{s.t.}
& & \sum\limits_{k \in \mathcal{S}} P_{t,k} \leq  P_c
\end{aligned}
\label{opt1p}
\end{eqnarray}
according to the channel capacity and
\begin{eqnarray}
(\textbf{P4}): \quad
\begin{aligned}
& \underset{\mathcal{P}_\textbf{t,q}}{\text{max}}
& & Q_T \\ 
& \text{s.t.}
& & \sum\limits_{k \in \mathcal{S}^c} P_{t,k} \leq  P_q
\end{aligned}
\label{opt2p}
\end{eqnarray}
according to the harvested power. Here, $\mathcal{P}_\textbf{t,c}$ and  $\mathcal{P}_\textbf{t,q}$ are the sets of transmit power values for $k \in \mathcal{S}$ and $k \in \mathcal{S}^c$, respectively. $P_c$ and $P_q$ represent the total transmit power allocated to information transfer subcarriers and power transfer subcarriers, respectively, $P_c=P_{t,e} \sum_{k=1}^K s_k$ and $P_q=P_{t,e} \sum_{k=1}^K s^{c}_{k}$. 

Both of $(\textbf{P3})$ and $(\textbf{P4})$ are convex optimization problems. In order to solve  (\textbf{P3}), we define the Lagrange function as 
\begin{eqnarray}
L(P_t, \lambda)=C_T + \lambda(P_c- \sum_{k \in \mathcal{S}} P_{t,k})
\label{lagran}
\end{eqnarray}
where $\lambda$ is the Lagrange multiplier. By differentiating the Lagrange function with respect to $P_{t,k}$, we obtain
\begin{eqnarray}
P_{t,k}= \left( \frac{B}{\lambda \ln2} - \frac{\sigma^2_z}{|H_k|^2}\right)^+, 
\label{opt1pP}
\end{eqnarray}
where $(x)^+=\max(0,x)$, and $\frac{B}{\lambda \ln2}$ is the water level.

The solution of the optimization problem $(\textbf{P4})$ is to allocate all transmit power to the subcarrier that ensures the maximum of $\eta_k |H_k|^2$ value. It is expressed as 
\begin{eqnarray}
P_{t,k}=
\begin{cases} 
P_q,& k= \text{arg} \: \underset{k}{\text{max}} (\eta_k |H_k|^2) \\ 0,& \text{otherwise}
\end{cases}
\end{eqnarray}
In case of limited transmit power value $(P_{t,max})$ for each individual subchannel, a new constraint needs to be added to $(\textbf{P4})$ as $P_{t,k} \leq P_{t,max}$, $\forall k$. $(\textbf{P4})$ becomes
\begin{eqnarray}
(\textbf{P5}): \quad
\begin{aligned}
& \underset{\mathcal{P}_\textbf{t,q}}{\text{max}}
& & Q_T \\ 
& \text{s.t.}
&& \sum\limits_{k \in \mathcal{S}^c} P_{t,k} \leq  P_q, \\
&&& P_{t,k} \leq P_{t,max}, \forall k.
\end{aligned}
\label{opt3p}
\end{eqnarray}
In that case, $(\textbf{P5})$ is a linear programming problem and solved by the interior point method. 

The subcarrier powers  allocated according to the PA problems can be used in the SA problems. Moreover, the values of decision variables calculated as the solutions of SA and PA problems can be recursively used for both optimization problems, to obtain a performance closer to the optimal solution. Alternatively, the joint subcarrier and power allocation can be considered as a future work, although it is possible to encounter higher computational complexity.

\begin{figure}[t]
\centering

\includegraphics[trim = 1.5cm 0.5cm 1.5cm 0cm, clip, width=0.99\linewidth]{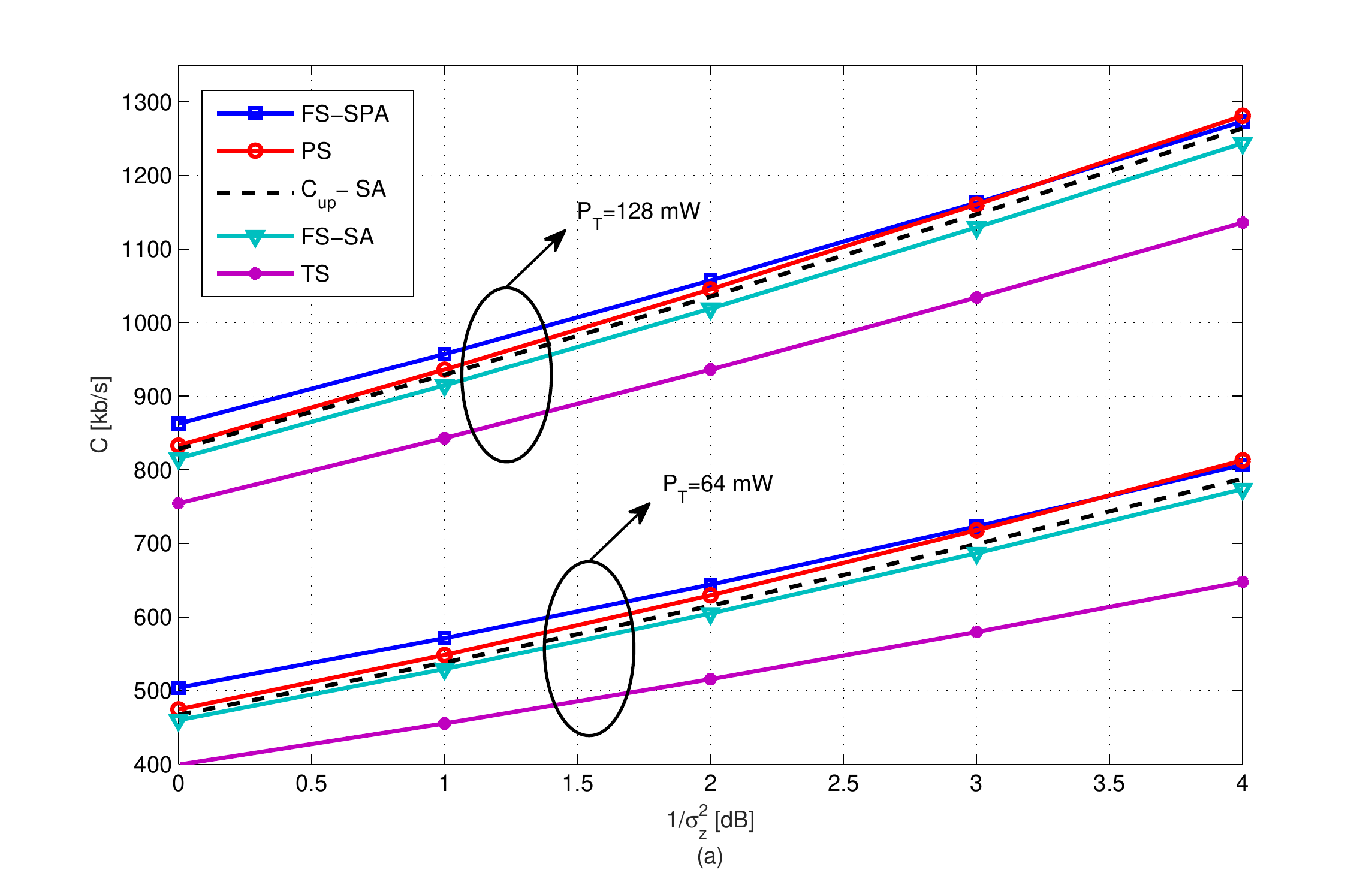}

\caption{Optimal values of channel capacity for TS, PS, FS-SA, FS-SPA and $C_{up}$ for SA ($\mathrm{C_{up}}$-SA) are plotted vs. $1/\sigma^2_z$ based on the optimization problems \textbf{(P1)} and  \textbf{(P3)}. $P_T = 64$ and $128$ mW.}
\label{sim1a}
\end{figure}

\vspace{0.2cm}
\section{Numerical Results}

Numerical studies are conducted to demonstrate the performance of the proposed SWIPT system. 
The results of optimization approaches are compared with TS and PS based receivers \cite{6489506}. Monte Carlo simulations are run for $10^6$ multi-carrier symbols with $K=32$ subcarriers. The parameters are selected as  $B=15$ kHz and $\eta_k=0.5$, $\forall k$. The total transmit powers are $P_T=K \times  P_{t,e}=64$ and $128$ mW. The channel coefficients are  Rayleigh distributed, path loss is not included.  

Firstly, we examine the results of (\textbf{P1}) and (\textbf{P3}). The minimum required power value is taken as $Q_{min}=12$ mW. For (\textbf{P1}), the transmit power for each subcarrier is constant, $P_{t,k}=P_{t,e}=2$ and $4$ mW, $\forall k$. The obtained subcarriers are used for (\textbf{P3}). The optimal values are calculated for SA of FS (FS-SA) and SPA of FS (FS-SPA) as well as both TS and PS based receivers. The channel capacity values versus the inverse of noise variance are shown in Fig. \ref{sim1a}. The total channel capacity increases, in line with the upper bound in (\ref{upperbound1}). From the results, it can be seen that at higher (lower) noise levels FS-SPA(PS) provides superior performance.

\begin{figure}[t]
\centering

\includegraphics[trim = 1.8cm 0.5cm 1.5cm 0cm, clip, width=0.99\linewidth]{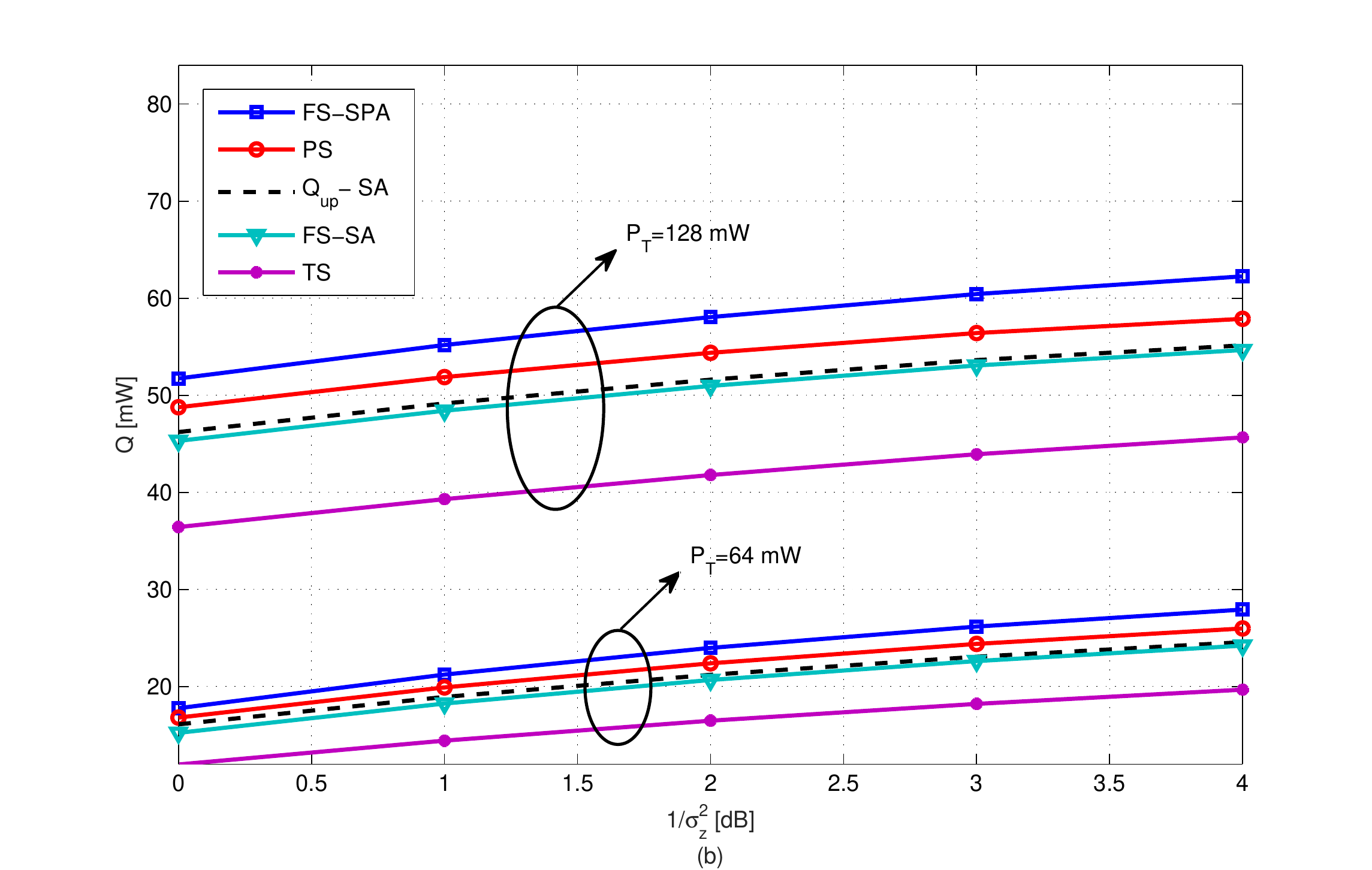}

\caption{Optimal values of harvested power for TS, PS, FS-SA, FS-SPA  and $Q_{up}$ for SA ($\mathrm{Q_{up}}$-SA) are plotted vs. $1/\sigma^2_z$ based on the optimization problems \textbf{(P2)} and \textbf{(P5)}. $P_T = 64$ and $128$ mW.}
\label{sim1b}
\end{figure}

\begin{figure}[t]
\centering

\includegraphics[trim = 2cm 1.5cm 1cm 0cm, clip, width=0.95\linewidth]{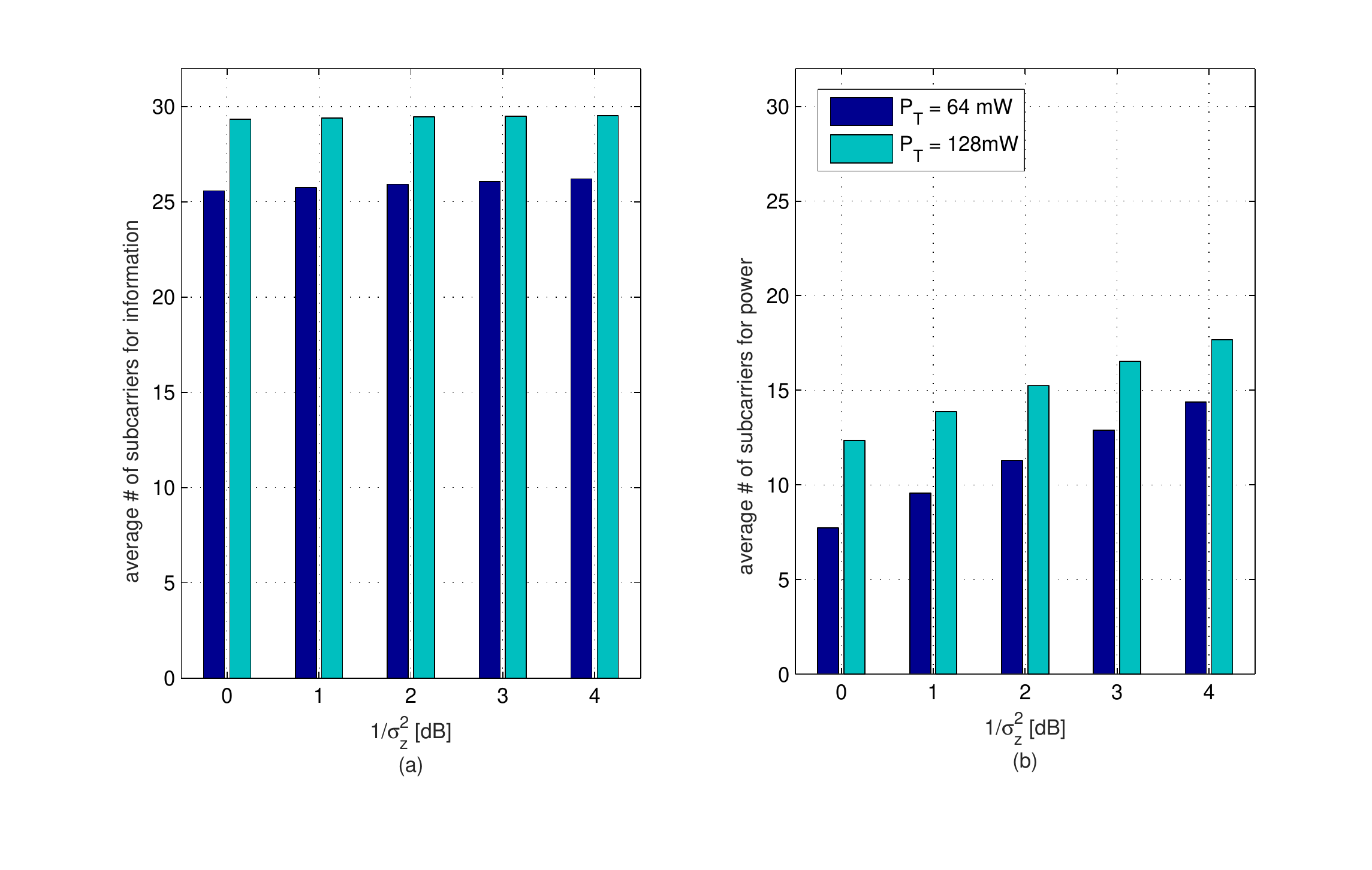}

\caption{Average numbers of subcarriers allocated for information transfer and power transfer are plotted vs. $1/\sigma^2_z$ for the optimization problems  \textbf{(P1)} in (a) and \textbf{(P2)} in (b), respectively. $P_T = 64$ and $128$ mW.}
\label{sim2}
\end{figure}

Considering (\textbf{P2}) with constant transmit power and (\textbf{P5}) with $P_{t,max}=2.4$, $4.8$ mW and $C_{min}=400$ kb$/$s, the harvested power values  shown in Fig. \ref{sim1b} are obtained. The harvested power increases with the increase of transmit power. FS-SPA outperforms all other considered techniques. The average numbers of subcarriers allocated to information transfer and power transfer are shown in Fig. \ref{sim2} (a) and (b), respectively.  For (\textbf{P1}), the average number of subcarriers allocated to information transfer increases slightly with the decrease of noise variance. For (\textbf{P2}), the average number of subcarriers allocated to power transfer rises noticeably with the reduced noise variance, as expected. 

\section{Conclusion}
\label{sec:conclusion}
In this paper, a new receiver structure is proposed to improve the performance of SWIPT in MCC systems. The frequency switch in the receiver forwards subcarriers to either the information decoder or the energy harvester. The optimal subcarrier selection approaches are developed to maximize the capacity or the harvested power using dynamic programming. In addition, the transmit power is optimized to increase the system performance. For the channel capacity, the proposed system model performs better than TS and PS based models particularly at high noise conditions. For the harvested power, the proposed model outperforms the existing approaches. 



\balance

\end{document}